\documentclass{aa}
\usepackage{graphicx}
\usepackage{txfonts}
\begin{document}

\title{Grid of theoretical NLTE equivalent widths of four  
\ion{Ba}{ii} lines and barium abundance in cool stars
\thanks{Tables 4 and 5 are only available in electronic form at the CDS
via the anonymous ftp to cdsarc.u-strasbg.fr (130.79.128.5)
or via http://cdsweb.u-strasbg.fr/cgi-bin/qcat?J/A+A/}
} 
\author 
{S. A. Korotin\inst{1}\and
S. M. Andrievsky\inst{1,2}\and
C. J. Hansen\inst{3}\and
E. Caffau\inst{2,4}\and
P. Bonifacio\inst{2,4}\and
M. Spite\inst{2}\and
F. Spite\inst{2}\and
P. Fran\c cois\inst{2}
}

\institute {
Department of Astronomy and Astronomical Observatory, Odessa National 
University, and Isaac Newton Institute of Chile Odessa branch,   
Shevchenko Park, 65014 Odessa, Ukraine\\
\email{serkor@skyline.od.ua}
\and
GEPI, Observatoire de Paris, PSL, Research University,  
CNRS,  Univ Paris Diderot, Sorbonne Paris Cit\'e, 
Place Jules Janssen, 92195 Meudon, France
\and
Dark Cosmology Centre, Niels Bohr Institute, The University of Copenhagen,
Juliane Maries Vej 30, 2100 Copenhagen \O, Denmark,
\and   
Istituto Nazionale di Astrofisica - Osservatorio Astronomico di Trieste,
via Tiepolo 11, 34143 Trieste, Italy} 

\date{}
\authorrunning{Korotin et al.}
\titlerunning{NLTE corrections}%
\abstract
{We present a grid of computed non-local thermodynamic equilibrium (NLTE) 
equivalent widths
(EW) and NLTE abundance corrections for four \ion{Ba}{ii} 
lines: 4554, 5853, 6141, and 6496~\AA.}
{The grid can be useful in deriving the NLTE barium abundance 
in stars having parameters in the following ranges:
effective temperature from 4000\, K to 6500\,K, surface gravity 
$\log~g$  from 0 to 5, 
microturbulent velocity 0 km~s$^{-1}$ to 3 km~s$^{-1}$, metallicity [Fe/H] 
from --2 to +0.5, and [Ba/Fe] from --0.4 to +0.6. The NLTE abundance 
can be either derived by EW interpolation (using 
the observed \ion{Ba}{ii} line EW) or by using the NLTE correction 
applied to a  previously  determined LTE abundance.}
{\ion{Ba}{ii} line equivalent widths and the NLTE corrections were 
calculated using the updated MULTI code and \ion{the Ba}{ii} atomic model 
that was previously applied to determine the NLTE barium abundance in 
different types of stars.}
{The grid is available on-line through the web, and we find that the grid Ba 
NLTE corrections are almost as accurate as direct NLTE profile fitting (to
within 0.05--0.08\,dex). For the weakest \ion{Ba}{ii} line 
(5853~\AA) the LTE abundances almost agree with the NLTE abundances, whereas the other three 
\ion{Ba}{ii} lines, 4554, 6141, and 6496~\AA,~ need NLTE corrections even at 
the highest 
metallicities tested here. The 4554~\AA\ line is extremely strong and should not be 
used for abundance analysis above [Fe/H]=$-1$. Furthermore, we tested the 
impact of different model atmospheres and spectrum synthesis codes and found
average differences of 0.06\,dex and 0.09\,dex, respectively, for all four 
lines. At these metallicities we find an average $\Delta$NLTE of $\pm0.1$\,dex
for the three useful Ba lines for subsolar cool dwarfs.}
{}

\keywords {Line: profiles -- Stars: abundances -- Stars: late-type}

\maketitle

\section{Introduction}

Barium  is one of the chemical elements that attracts close attention
of the specialists working in the field of the Galactic chemical evolution. 
This element is mainly produced by the $s$-process in low-mass AGB stars (e.g., \citealt{Bus99}; \citealt{Cri11}; \citealt{Bis10};
\citealt{Kar14}). It serves as a reference element in an 
estimation of the relative efficiency of the $s$- and $r$-process 
nucleosyntheses (see e.g. \citealt{Chr04}; \citealt{Han12}).

Usually, the four strongest lines of ionized barium in the visual part 
of the spectrum are studied. They are located at 4554, 5853, 6141, and 6496~\AA.
These 
lines can be quite strong in the spectra of solar metallicity
stars, and therefore they require the NLTE consideration. 
On the other hand, in metal-poor stars they can have much smaller 
equivalent widths, which may seem to be appropriate for the LTE analysis 
from an equivalent width perspective. Nevertheless, in such stars 
the concentration of free electrons in the atmosphere  is decreased, which 
causes a significant deviation from LTE. To derive the correct  
barium abundance in stars of different metallicities, one therefore
needs to 
perform either the direct barium line profile fitting (the NLTE synthetic 
spectrum technique, which is the best way), or to take into account
the calculated NLTE corrections in an LTE analysis (the less reliable way). 
Despite the lower reliability, the second approach can be very effective 
for estimating the barium abundances in large samples of stars. 

This paper presents a grid of NLTE barium line equivalent 
widths and NLTE corrections in cool stars that can be used for 
the barium abundance estimate. The method of the NLTE calculations 
is described in the next section. 

\section{\ion{Ba}{ii} atomic model and NLTE calculations}

Our \ion{Ba}{ii} atomic model was first described among other things 
in \citet{And09}, where it was applied to investigate a 
sample of very metal-poor Galactic halo stars (cool dwarfs and giants). 
It was later used to study Galactic thick- and thin-disc intermediate-to-cool 
dwarfs of different metallicities (\citealt{Kor10}, \citealt{Kor11}, 
\citealt{Mis12}, \citealt{Mis13c}), large proper motion metal-deficient 
G-K stars \citep{Klo11a,Klo11b}, red giants in the globular cluster 
NGC~6752 (\citealt{Dob12}, \citealt{Thy14}), red giants of the globular 
clusters M~10 and M~71 \citep{Mis09}, cool semiregular variable giants of 
the SRd type \citep{Bri10, Bri12}, RR~Lyr-type stars (\citealt{And10}, 
\citealt{Han11}), intermediate-to-cool dwarfs in open clusters 
(\citealt{Dor12}, \citealt{Mis13a,Mis13b,Mis14a,Mis14b}), 
and intermediate-to-cool supergiant stars of  nearly solar metallicity 
in the Galactic thin-disc stars \citep{And13, And14}.

Below we briefly describe our model of the barium atom. It consists of 31 levels 
of \ion{Ba}{i}, 73 levels of \ion{Ba}{ii}, and the ground level of 
\ion{Ba}{iii}. Ninety-one bound-bound transitions between the first 28 levels 
of \ion{Ba}{ii} (n < 12 and l < 5) are computed in detail. Populations of the 
remaining levels are used to conserve the particle number. For two levels,
5d${}^2$D and 6p${}^2$P$_{0}$, the fine structure is taken into account. The 
information about the adopted oscillator strengths, photoionization cross-sections, 
collisional rates, and broadening parameters can be found in \citet{And09}. 
Atomic level populations were determined using the MULTI code of 
\citet{Car86} with modifications as given in \citet{Kor99}. The 
MULTI code calculates the line profile for each line that is considered in 
detail. The line profiles are computed assuming NLTE approximation 
and depend upon many parameters: the effective temperature of the model, the 
surface gravity, the microturbulent velocity, and the line damping, as well 
as the populations of the relevant levels. 

The barium atom has seven isotopes. For the two odd isotopes ${}^{135}$Ba and 
${}^{137}$Ba the hyper-fine structure is important. This effect is  most 
pronounced for the \ion{Ba}{ii} lines 4554 and 6496 \AA. As  was 
shown by \citet{Mas99} for adequate barium line 
modelling, it is sufficient to use the three-component model suggested by 
\citet{Rut78}. To calculate the Ba line profiles in the spectra 
of young stars, the even-to-odd abundance ratio of 
82:18 can be used as a rule \citep{Cam82}, and for the very old stars the ratio is 50:50. For our grid 
we used the ratio 82:18.

All the atmosphere models used for the grid of NLTE equivalent widths 
and correction calculations were interpolated basing on the ATLAS9 
model atmosphere grid of \citet{CK03}.

\section{Grid of the NLTE equivalent widths and
corrections \label{sec:grid}}

The following stellar parameter ranges are covered by our grid,
which focuses on more metal-rich stars (compared to the very metal-poor
and extremely metal-poor stas) that are typically targeted in current and 
future surveys:

-- effective temperature: 4000 - 6500* K, step = 250 K;

-- surface gravity: 0 - 5, step = 0.5

-- microturbulent velocity: 0 - 3 km~s$^{-1}$, step = 1 km~s$^{-1}$;

-- metallicity: [Fe/H] = +0.5, 0.0, --0.5, --1.0, --1.5 and --2.0;

-- relative barium abundance: [Ba/Fe] = --0.40, --0.20, 0.00, +0.20,
+0.40, +0.60.

For the models with [Fe/H] below --1.00 we calculated NLTE equivalent 
widths with an increased atmosphere abundance of $\alpha$-elements 
($\alpha$/Fe] = +0.4), while for a metallicity of --0.5 both cases
(solar alpha-element abundance and an increased one) were considered. 
The asterisk indicates that at this temperature we do not have NLTE 
calculations for $\log~g = 0$, since there is no existing model atmosphere 
for this combination of stellar parameters.

The NLTE equivalent widths of the four barium lines were calculated:
4554, 5853, 6141, and 6496 \AA. The corresponding parameters of these
lines are given in Table \ref{hfs}.

\begin {table}
\caption {Barium line parameters.}
\label {hfs}
\begin{tabular}{l c c c c }
\hline
$\rm \lambda (\AA)$ & HFS & \it f & $\rm \log~\gamma_{rad}$ & $\log~ 
\Gamma_{\rm{vw}}$\\
\hline
             &$\Delta\lambda$, m\AA& &&  solar   \\
\hline
4554.03&   0&  0.597&  $ 8.20    $&  -7.60\\
       &  18&  0.081& \\
       & -34&  0.049& \\
\\
5853.70&   -&  0.025&  $ 8.20    $&  -7.47\\
\\
6141.71&   -&  0.140&  $ 8.20    $&  -7.47\\
\\
6496.90&   0&  0.086&  $ 8.10    $&  -7.47\\
       &  -4&  0.012&\\
       &   9&  0.007&\\ 
\hline  
\end {tabular}  
\end {table}   

Tables 4 and 5 (available in electronic form at the CDS) contain the NLTE 
equivalent widths and NLTE corrections, respectively. For each barium 
line we selected six values of [Ba/Fe] (--0.4, --0.2, 0.0, +0.2, +0.4 
and +0.6) for the NLTE EW grid, and three values of [Ba/Fe] 
(--0.2, 0.1, and +0.4) for the NLTE correction grid. For each 
of these values we list in the corresponding table the EWs or corrections 
calculated for the full set of  effective temperature, 
surface gravity, microturbulent velocity, and metallicity. 

The grid is also available at the web-site  
{\tt http://nlte.obspm.fr} 
together with an interpolation program. The latter enables deriving 
the absolute barium abundances, namely 12.00 + $\log \epsilon$(Ba), by 
inserting the stellar atmosphere parameters, specifying the line of 
interest, and its measured equivalent width. If a star has a very 
anomalous strength of a barium line for the adopted parameters, that is, an 
anomalous barium abundance that is either lower than [Ba/Fe] = --0.4, 
or higher than [Ba/Fe] = +0.6, then the program gives a notification 
that the expected barium abundance is out of the grid range. In this 
case, the better way to derive the barium abundance is to apply a direct 
profile-fitting procedure.

\section{Comparing NLTE abundances based on the grid and profile fitting}

The most reliable way to derive the NLTE abundance from a given line is 
to compute an NLTE profile synthesis and fit the 
observed profile. Nevertheless, this procedure may become complicated when 
a large amount of stars are to be investigated. A faster, and therefore more 
efficient, method for analysing large samples of stars is to determine the 
NLTE abundances by interpolating within a grid of NLTE equivalent widths. 
Such a grid was presented in Sect.~\ref{sec:grid}. Below we give the results 
of our grid tests and show which differences between the profile-based and 
grid-based NLTE abundances can be expected for the stars with different sets 
of atmosphere parameters and different barium lines.

\begin{figure}
\includegraphics[width=8.5cm] {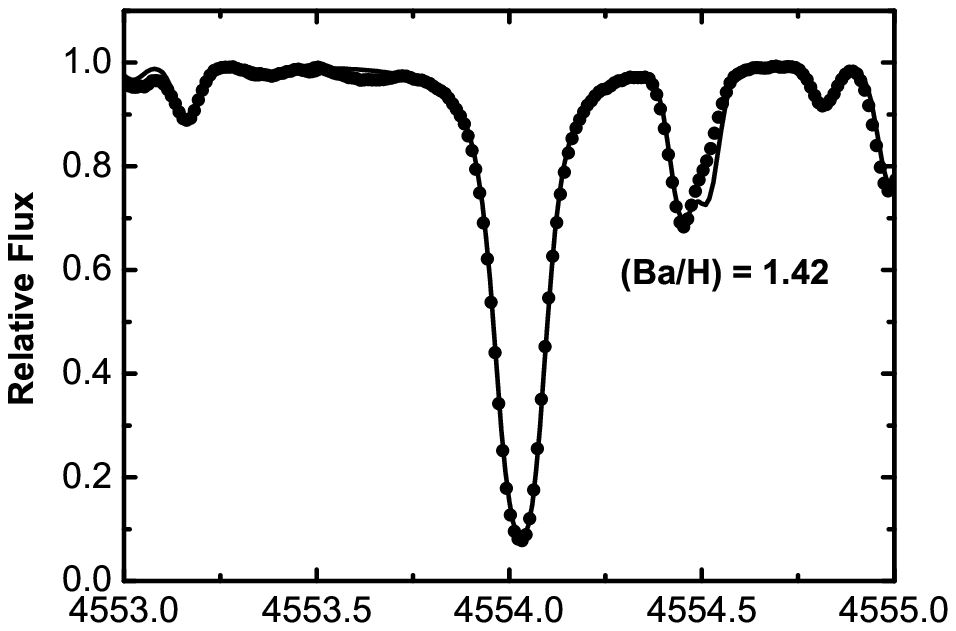}
\includegraphics[width=8.5cm] {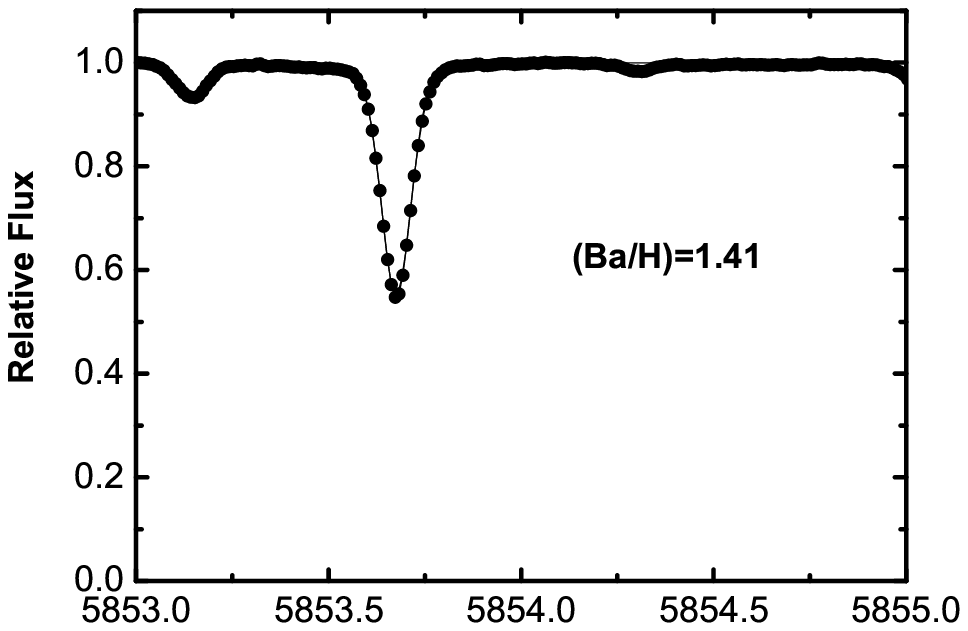}
\includegraphics[width=8.5cm] {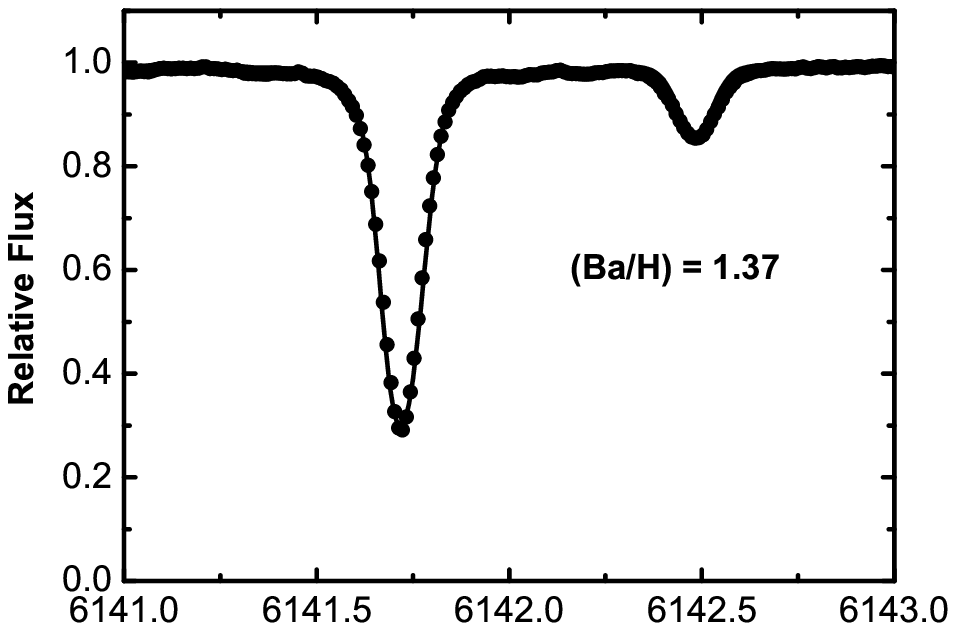}
\includegraphics[width=8.5cm] {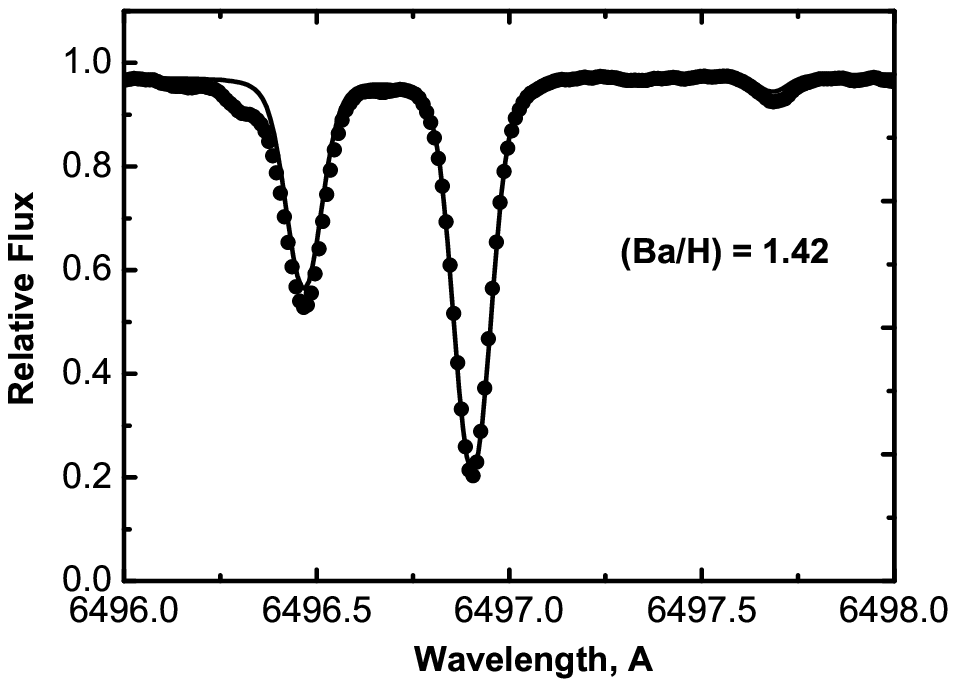}
\caption[]{Best NLTE fits for the barium line profiles in the
spectrum of HD010700. Dots denote the observed spectrum, the
continuous line the calculated 
spectrum. The absolute NLTE barium abundances (i.e. $\log \epsilon(\rm Ba)
+12.00$) derived from individual lines are indicated in each plot.}
\label {prof}
\end{figure}

\begin{figure}
\resizebox{\hsize}{!}                   
{\includegraphics {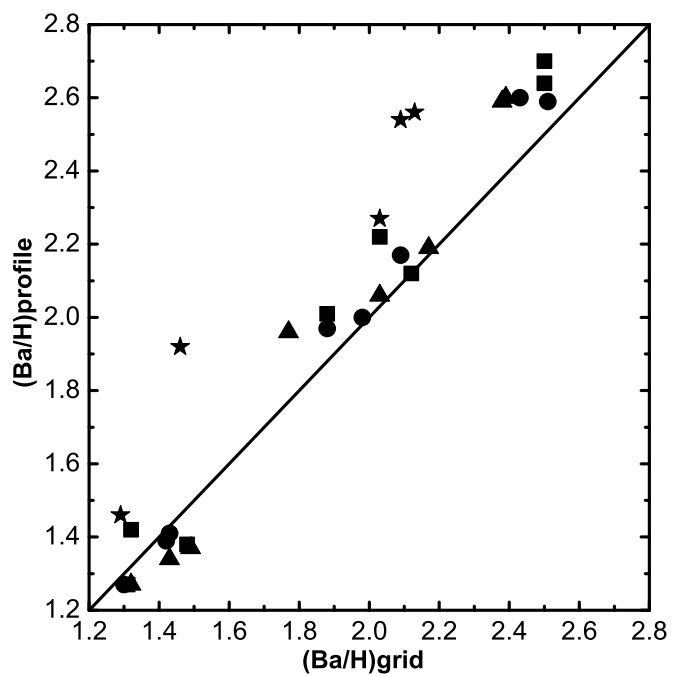}}
\caption[]{Graphical comparison of the NLTE barium profile-based and 
grid-based abundances derived from the four lines (asterisks - 4554 \AA~, 
circles - 5853 \AA~, triangles - 6141 \AA~, squares - 6496 \AA) for 
eight test stars.}
\label {comp}
\end{figure}

We have selected eight test stars (see Table \ref{tab:nlteab}). The spectra of 
HD stars and their atmospheric parameters are described in \citet{Jof14}. 
Similar data for VW Dra can be found in \citet{Bri12}. For these stars we 
performed the direct NLTE synthesis of the barium line profiles (an example 
for one star is shown in Fig. \ref{prof}), and then, 
using the measured equivalent widths and corresponding atmospheric 
parameters, we derived the grid-based NLTE barium abundances. The results 
are presented in Table \ref{tab:nlteab} and are shown in Fig. \ref{comp}. The 4554\,\AA~ line is very strong (even in the intermediate 
metal-poor stars). In the measurements we therefore underestimate its 
equivalent widths (far wings are missed). As a result, our grid-based NLTE 
abundances appear to be systematically lower than abundances derived from 
the direct profile synthesis. This means that the 4554\,\AA~ line has to be 
excluded from the consideration because it is not appropriate for the grid-based 
abundance determination (there is the same systematic trend for other
lines, although of lesser  magnitude). However, this is not a great loss for the future studies in the context of the Gaia-ESO Survey \citep{Gil12}, for example, 
since this line is not covered by the standard survey wavelength 
setup, and excluding this line will therefore not have a negative 
impact on the final Ba abundance. Furthermore, the difference between 
the direct NLTE line profile fitting and the NLTE grid-based abundances for 
the 4554\,\AA~ line can be as large as $\pm0.46$\,dex (on average 
$\pm 0.29$\,dex). This is larger than the desired  abundance 
uncertainties. This implies that the 4554\,\AA~ line should indeed not 
be used for accurate abundance measurements in more metal-rich stars 
([Fe/H]$\ga -1$). Nevertheless, in faint star spectra that have a low 
S/N ratio, only a handful of strong lines are measurable. In this case, even the 4554\,\AA~ line may provide (with a corresponding 
correction) some rough estimate of the barium content. 

The situation with the other three lines is much more promising. These 
lines are not too strong, even in the spectra of stars with solar
metallicity. In particular, they are weak enough in the metal-poor stars, 
and this guarantees quite reliable results on barium  NLTE abundances derived 
from the grid. We find an average difference of $< \pm 0.05, 0.05, 0.08$\,dex 
for the 5853, 6141, and 6496\,\AA~ line, respectively, between the NLTE 
profile fitting and the grid Ba abundances. This small difference confirms 
that the grid abundances are very accurate for these three lines.

\begin{table*}
\begin{center}
\caption{Stellar parameters and NLTE barium abundances (grid- and 
profile-based) for the sample stars. 
\label{tab:nlteab}}
\begin{tabular}{lcccc|cccc|cccc}
\hline\hline
\multicolumn{1}{c}{Star} & \multicolumn{4}{c}{Stellar parameters} & 
\multicolumn{4}{r}{[Ba/Fe]$_{\rm grid}$ }& 
\multicolumn{4}{r}{[Ba/Fe]$_{\rm profile}$}\\
\hline
Ident.  & T$_{\rm eff}$ & $\log~g$ & [Fe/H] & $\xi$  & 4554 & 5853 & 6141 & 
6496 & 4554 & 5853 & 6141 & 6496 \\
            & [K] &[$g$ in cm~s$^{-2}$] &  & [km~s$^{-1}$]  &  &  &  &  &  &  &  &  \\
\hline
\object{HD010700}  & 5331 & 4.44 & $-0.53$& 1.1 &$-0.35$&$ -0.21$&$ -0.15$&$ -0.32$&$ -0.18$&$ -0.23$&$ -0.27$&$ -0.22$\\
\object{HD018907}  & 5069 & 3.45 & $-0.62$& 1.2 &  out  &$ -0.25$&$ -0.23$&$ -0.24$&$ -0.23$&$ -0.28$&$ -0.28$&$ -0.28$\\
\object{HD022049}  & 5050 & 4.60 & $-0.07$& 1.1 &$-0.07$&$ -0.01$&$ +0.07$&$ -0.07$&$ +0.17$&$ +0.07$&$ +0.09$&$ +0.12$\\
\object{HD100407}  & 5044 & 2.87 & $+0.21$& 1.1 &$-0.25$&$ +0.05$&$ +0.01$&$ +0.12$&$ +0.18$&$ +0.22$&$ +0.22$&$ +0.26$\\
\object{HD107328}  & 4590 & 2.20 & $-0.30$& 1.2 &$-0.41$&$ +0.01$&$ -0.10$&$ +0.01$&$ +0.05$&$ +0.10$&$ +0.09$&$ +0.14$\\
\object{HD113226}  & 4983 & 2.77 & $+0.12$& 1.1 &$-0.20$&$ +0.22$&$ +0.09$&$ +0.21$&$ +0.25$&$ +0.30$&$ +0.30$&$ +0.41$\\   
\object{HD220009}  & 4266 & 1.43 & $-0.67$& 1.3 &  out  &$ -0.08$&$ -0.07$&$ -0.02$&$ -0.12$&$ -0.11$&$ -0.16$&$ -0.12$\\ 
\object{VW Dra}    & 4660 & 2.00 & $+0.00$& 1.4 &   no  &$ -0.19$&$ -0.14$&$ -0.05$&    no  &$ -0.17$&$ -0.11$&$ -0.05$\\
\hline
\hline
\end{tabular}
\end{center}
\tablefoot{
\begin{itemize}
\item[2-6] Columns 2-5 from Jofre et al. (2014) - their Table 1 and 
references therein.
\item[2-6] out - out of range
\item[2-6] no - no line is available 
\end{itemize}}
\end{table*}

\section{LTE versus NLTE abundances}

Here we present the results of an independent LTE analysis with an NLTE
analysis based on a combined use of MULTI and ATLAS9 for the
seven stellar 
spectra described in the previous section.
For comparison, the LTE abundances were derived using interpolated 
MARCS model atmospheres \citep{Gus08} and the MOOG spectrum synthesis 
code (\citealt{Sne73}, version 2014). 
As  mentioned above, the stellar parameters and v$~sini$ were 
adopted from \citet{Jof14}, and the solar Ba abundance of 2.17 
(our NLTE value) was used for this study.  The line list compiled for 
the LTE analysis consists of Ba data from \citet{Gal12}, molecular data from 
\citet{Sne14}, and most of the remaining atomic data were adopted from the 
Kurucz database \footnote{http://kurucz.harvard.edu/linelists.html}. 
All LTE abundances were determined using line synthesis and the 
interactive minimum $\chi^2$ technique in MOOG. The stellar rotation and LTE 
abundances are listed in Table \ref{tab:abundance}.

\begin{table}
\begin{center}
\caption{Stellar parameters and abundances (LTE and NLTE) from direct 
profile fitting for the sample stars. 
\label{tab:abundance}}
\begin{tabular}{lc|ccccc}
\hline\hline
\multicolumn{1}{c}{Star} & \multicolumn{1}{c}{} 
& \multicolumn{4}{c}{[Ba/Fe]$_{\rm LTE}$ }\\
\hline
Identifier  &  v$~sini$  & 4554 & 5853 & 6141 & 6496  \\
            & [km~s$^{-1}$] &  &  &  &   \\
\hline
\object{HD010700} &  1.1    & $-0.08$     &$-0.18$ &$-0.06^u$ &$-0.15$\\
\object{HD018907} & $2.5^c$ & $-0.05$     &$-0.23$ &$-0.14$   &$-0.14$\\
\object{HD022049} &  2.4    & $+0.16$     &$+0.06$ &$+0.19$   &$+0.15$\\
\object{HD100407} &  2.4    & $+0.16^s$   &$+0.28$ &$+0.31$   &$+0.43$\\
\object{HD107328} &  1.9    & $+0.0^{u,s}$&$+0.2^u$&$+0.20$   &$+0.3^u$\\
\object{HD113226} &  2.0    & $+0.7^{u,s}$&$+0.43$ &$+0.34$   &$+0.44$\\   
\object{HD220009} &  1.0    &$-0.10^{u,s}$&$-0.09$ &$-0.01$   &$-0.01$\\ 
\object{VW Dra}   &  2.3    &  -- & --& --& --\\
\hline
\hline
\end{tabular}
\end{center}
\tablefoot{
\begin{itemize}
\item[c] $^{c}$ differs from Jofre et al. (2014) and references therein
\item[u] $^u$ uncertain values, poor LTE fit
\item[s] $^s$ saturated line
\end{itemize}}
\end{table}

\begin{figure}
\resizebox{\hsize}{!}                   
{\includegraphics {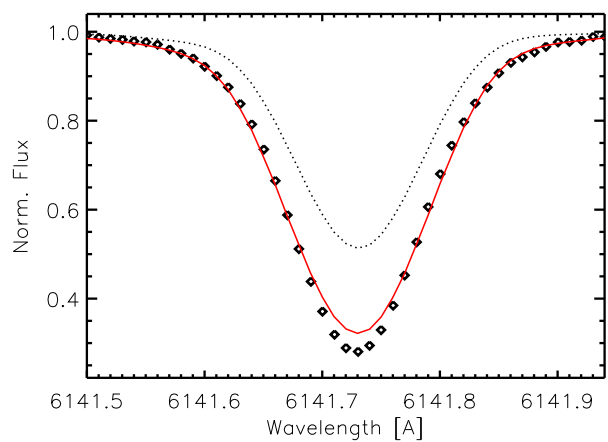}}
\caption[]{Barium line in HD18907 at 6141\,\AA~ fit in LTE (where [Ba/Fe] 
$= -0.14$ is shown as the red, solid line). The core is poorly fit, and the 
Fe blend is illustrated by the black, dashed line (synthesized with 
no barium).}
\label {LTE6141}
\end{figure}

Overall, the LTE and NLTE abundances agree within $\approx 0.1$\,dex for these 
stars (dwarfs and giants of around solar metallicity). The 6496\,\AA~ line 
yields fairly consistent abundances in both LTE and NLTE, when the 
metallicity is subsolar. Above solar metallicity, larger deviations can 
occur. The largest difference between the LTE and NLTE abundances is found 
for the \ion{Ba}{ii} line at 6141\,\AA, which is a blended line  
(see Fig.~\ref{LTE6141}). 

\begin{figure}
\resizebox{\hsize}{!}                   
{\includegraphics {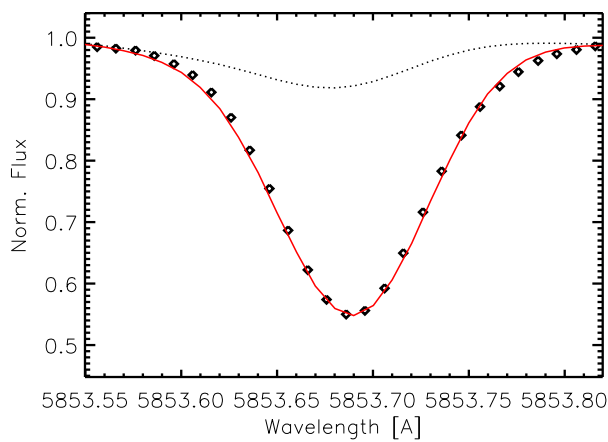}}
\caption[]{Barium line in HD10700 at 5853\,\AA~ well fit in LTE (the red, 
solid line has a [Ba/Fe] $= -0.18$). A weak blend is shown by the black, 
dashed line (synthesized with no barium).}
\label {LTE5853}
\end{figure}

A \ion{Fe}{i} line blends into the core of this Ba line. Different values for 
the $\log~gf$ of the blending \ion{Fe}{i} line translates into small changes 
in the Ba abundance of the 6141\,\AA~ line, and by increasing the 
$\log~gf$ of the Fe line by 0.15\,dex, the Ba abundance is reduced 
by 0.02\,dex. As seen from Figs.~\ref{prof}--\ref{comp}, we obtain good 
fits for the 6141 Ba line using the NLTE grid. The smallest difference 
between the LTE and NLTE abundances is found for the weakest Ba line, 
which is the 5853\,\AA~ line (see Fig.~\ref{LTE5853}). The blue 4554\,\AA~ 
line is saturated or even damped for most of these metal-rich stars and 
can be affected by uncertainties in the continuum placement (of the order 
of 0.03 to 0.07\,dex in the final  Ba abundance). For the red Ba 
lines the uncertainty from the continuum placement is smaller 
($\pm 0.01-0.03$\,dex) in these high-quality spectra.

The aforementioned differences in Ba abundance obtained independently using 
MULTI and ATLAS9 models versus MOOG and MARCS hide differences in the synthetic 
spectrum code, model atmosphere, atomic data, and continuum placement. 
Therefore, it is not the best ground to compare the LTE and NLTE abundances 
 to assess the size of the $\Delta$NLTE (NLTE$-$LTE abundance). 
However, the comparison gives a feeling for how abundance results 
of two independent studies obtained using completely different tools and 
methods -- only sharing the same high-resolution spectra - can differ. 
To determine which fraction of the reported $\approx 0.1$\,dex 
in Ba abundance is found between MULTI+ATLAS9 and MOOG+MARCS, we analysed the 
same star using MOOG+ATLAS9 to assess the abundance difference originating 
on one hand in the differing model atmospheres and on the other hand in the codes. The average [Ba/Fe] difference between using MOOG+MARCS and 
MOOG+ATLAS9 is 0.06\,dex for all four lines, and the largest difference is 
found for the 4554\,\AA~line. Meanwhile, the difference in codes result in 
an average difference of 0.09\,dex for all four lines. By adding these values 
in quadrature, obtain a total abundance difference of 0.11\,dex, 
which agrees with the aforementioned difference. We note that 
differences in the continuum placement were not taken into account, 
and the individual (atmosphere vs code) differences may therefore only 
be $\approx 0.05$\,dex. 

\begin{figure*}
\resizebox{\hsize}{!}                   
{\includegraphics {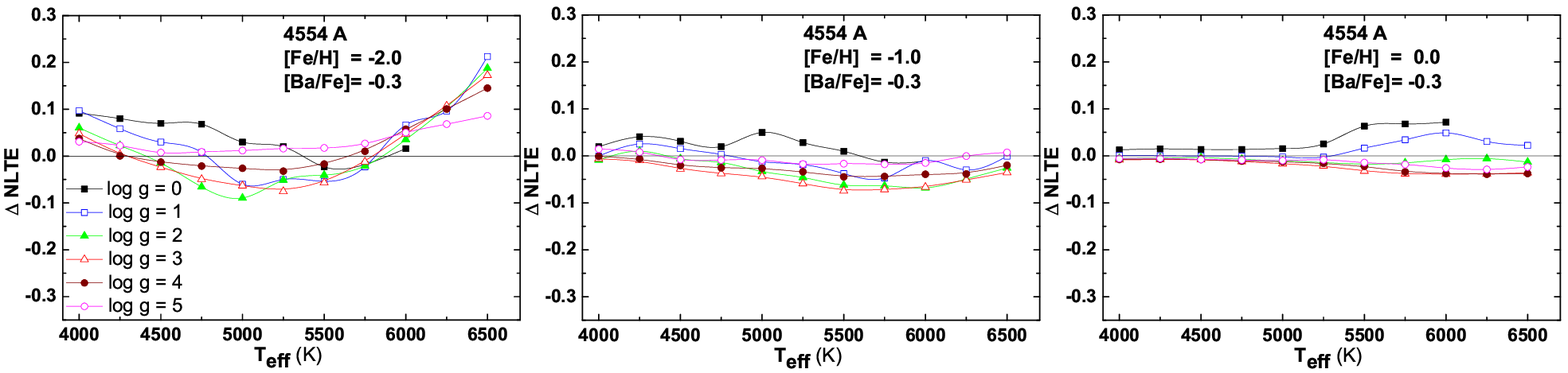}}
\resizebox{\hsize}{!}                   
{\includegraphics {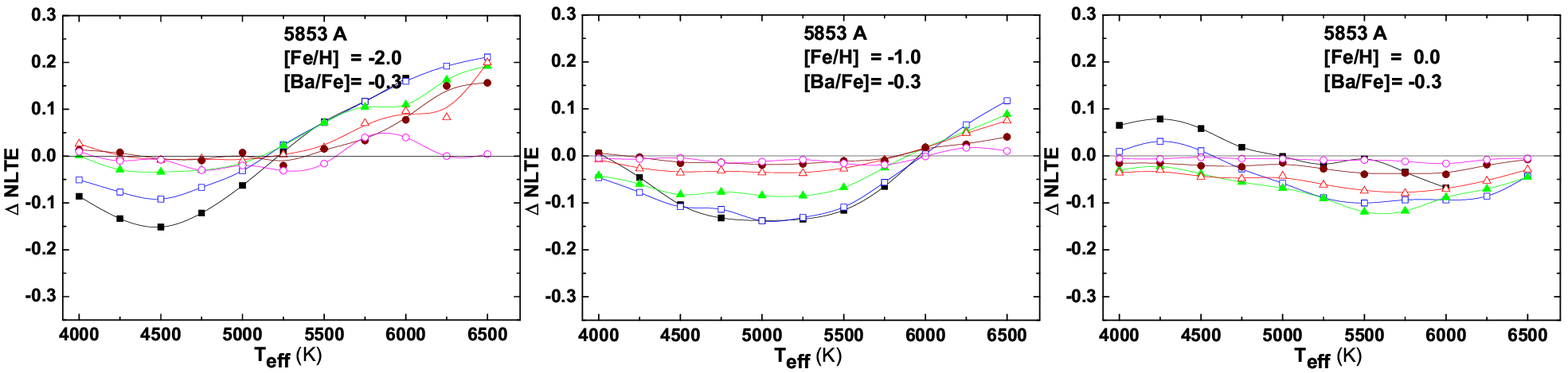}}
\resizebox{\hsize}{!}                   
{\includegraphics {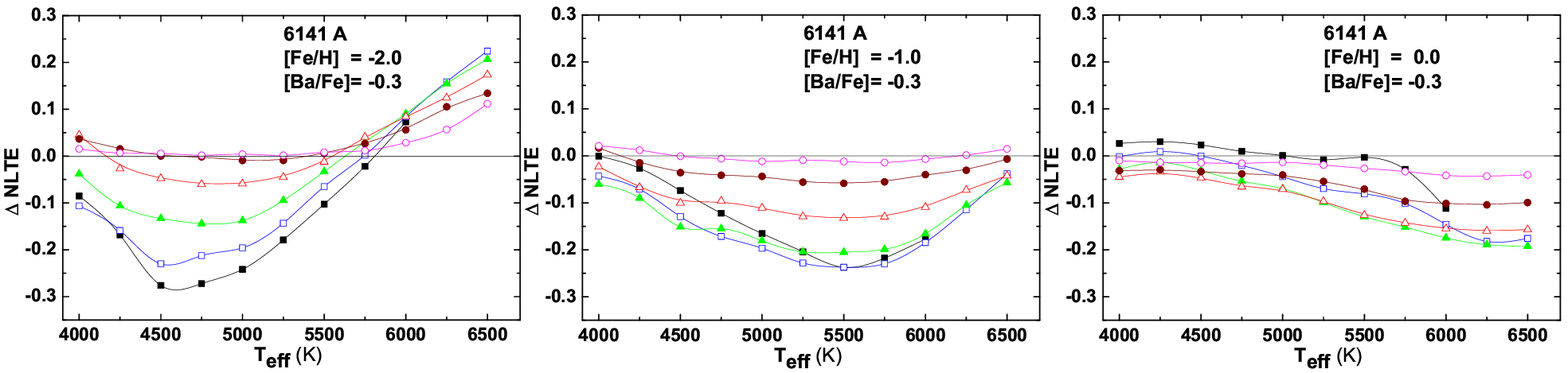}}
\resizebox{\hsize}{!}                   
{\includegraphics {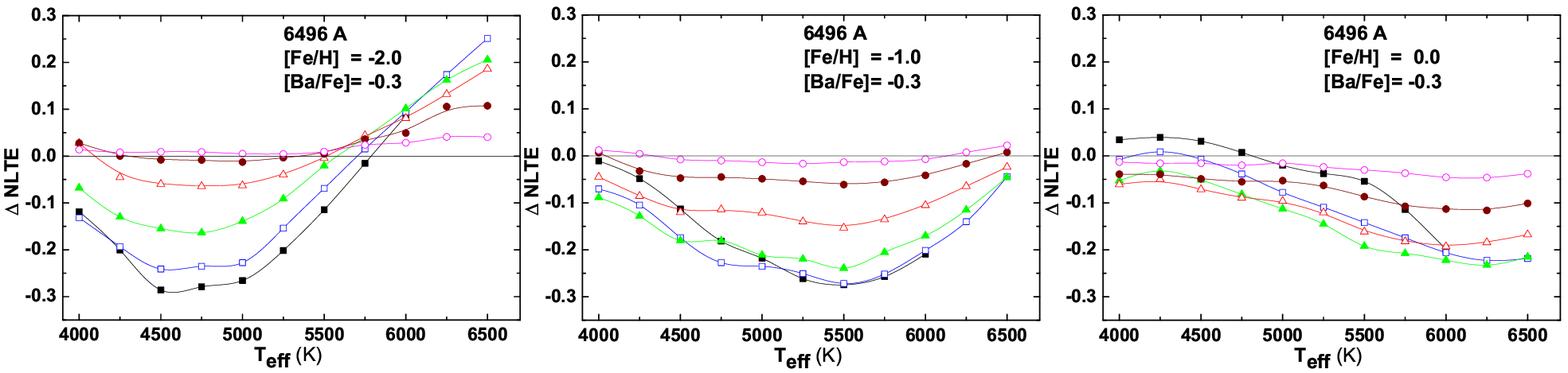}}
\caption[]{NLTE corrections ($\Delta$(NLTE) = NLTE -- LTE)
for the four barium lines as a function of 
effective temperature for the different gravities, metallicities, and
values [Ba/Fe] = --0.3. The gravity values are listed in the lower left 
corner of the upper plot. The adopted microturbulent velocity is 2 km~s$^{-1}$.}
\label {dnltem3}
\end{figure*}

\begin{figure*}
\resizebox{\hsize}{!}                   
{\includegraphics {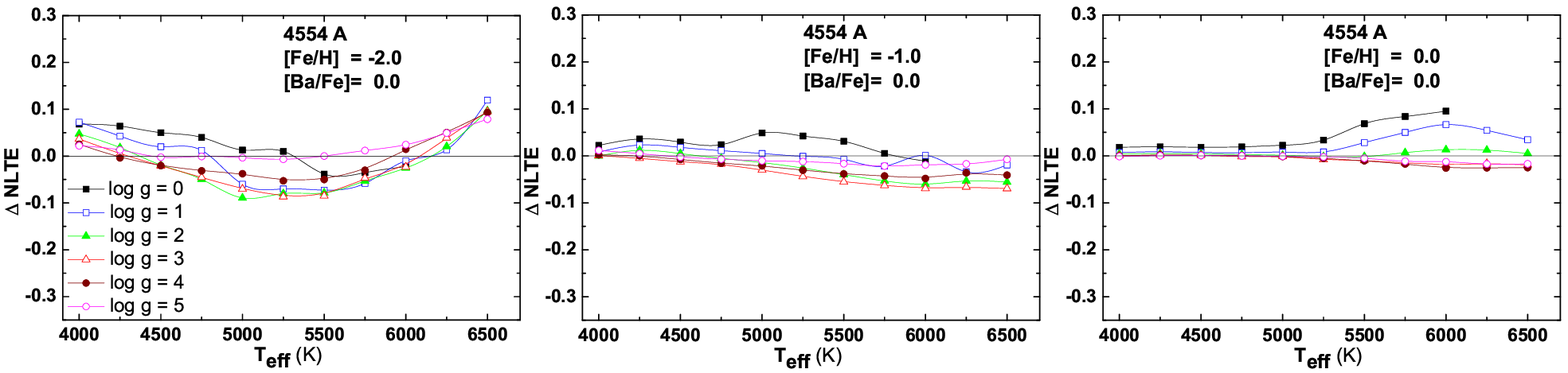}}
\resizebox{\hsize}{!}                   
{\includegraphics {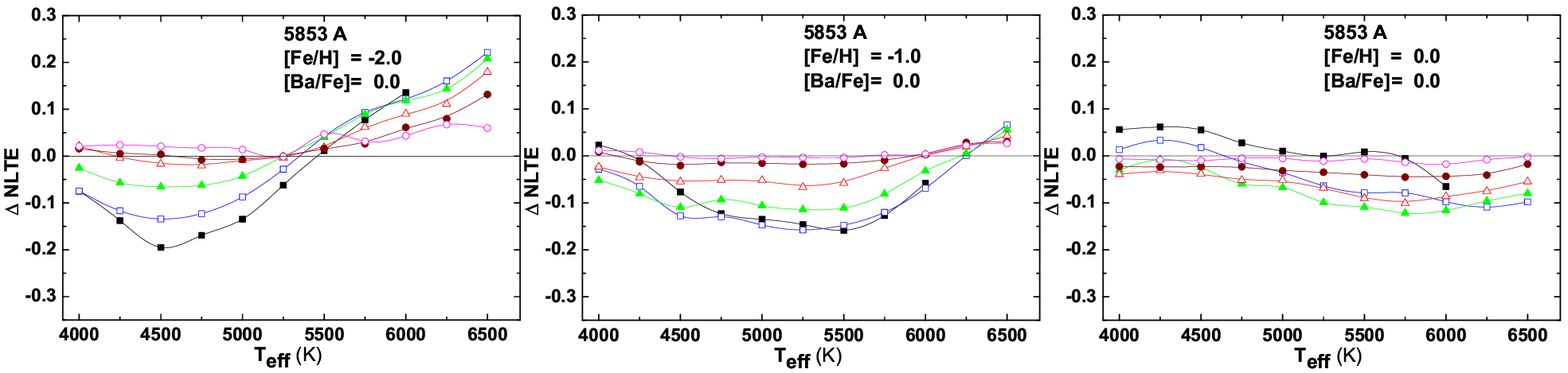}}
\resizebox{\hsize}{!}                   
{\includegraphics {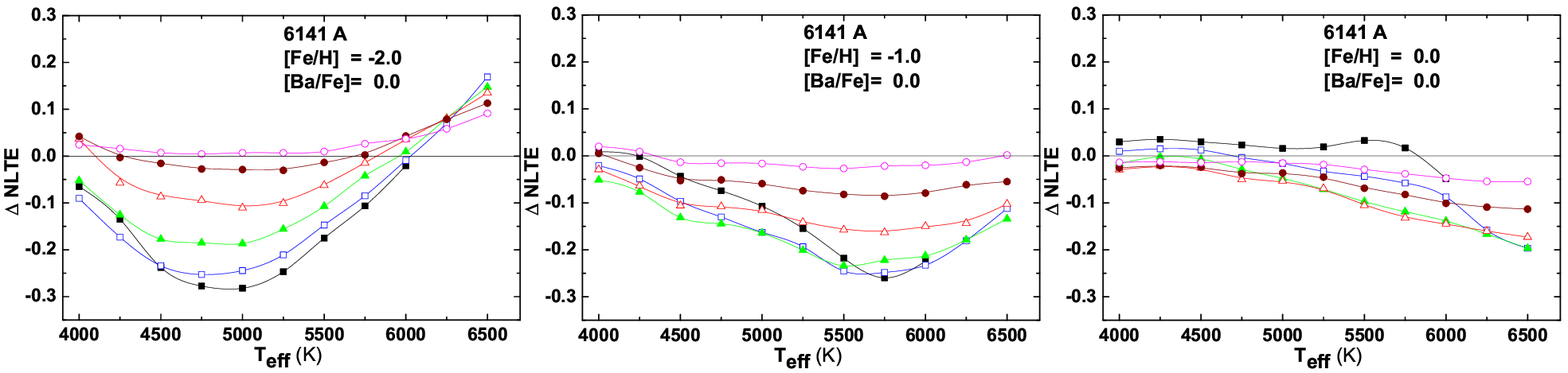}}
\resizebox{\hsize}{!}                   
{\includegraphics {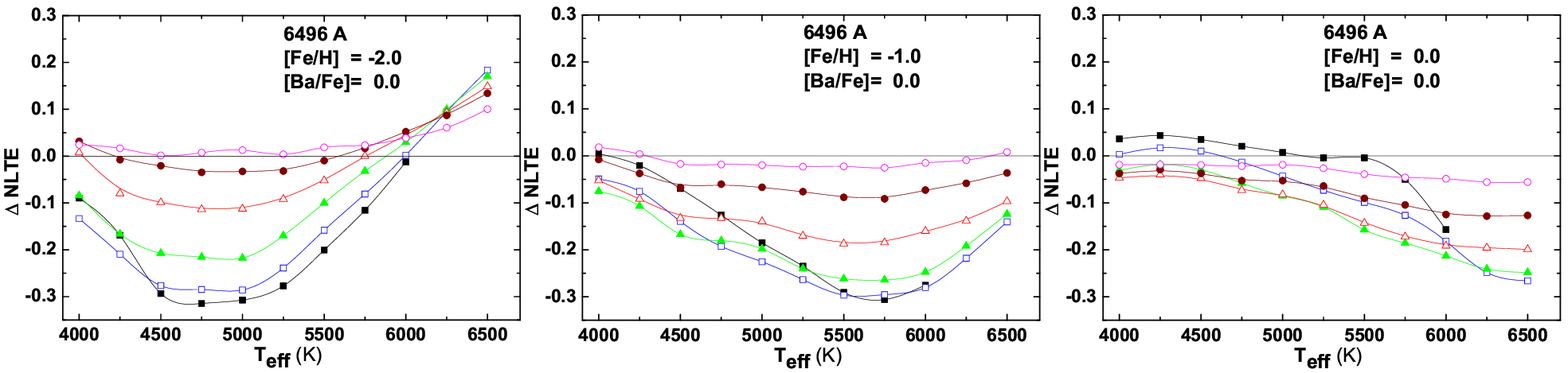}}
\caption[]{Same as Fig. \ref{dnltem3} for [Ba/Fe] = 0.}
\label {dnltep0}
\end{figure*}

\begin{figure*}
\resizebox{\hsize}{!}                   
{\includegraphics {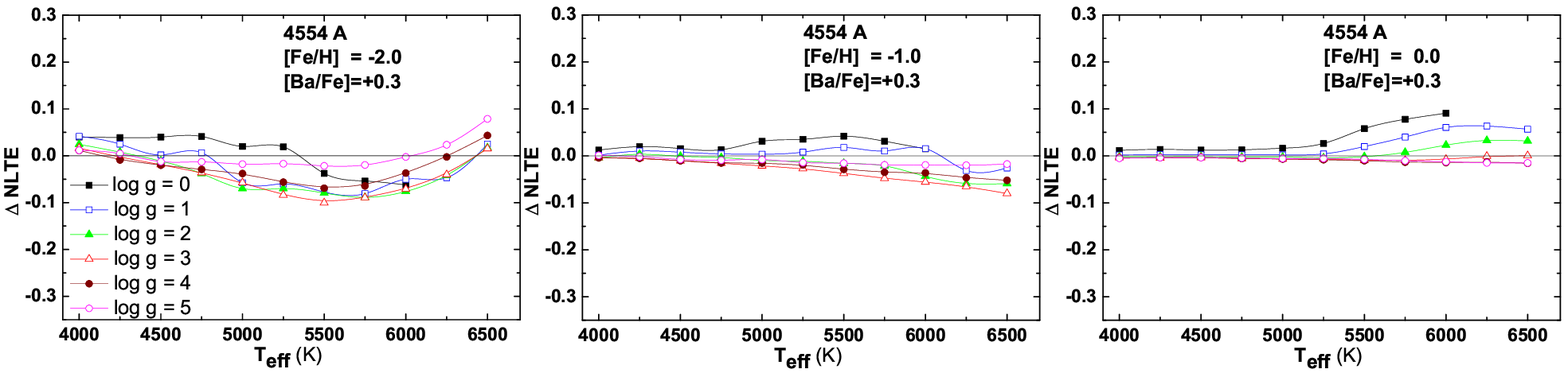}}
\resizebox{\hsize}{!}                   
{\includegraphics {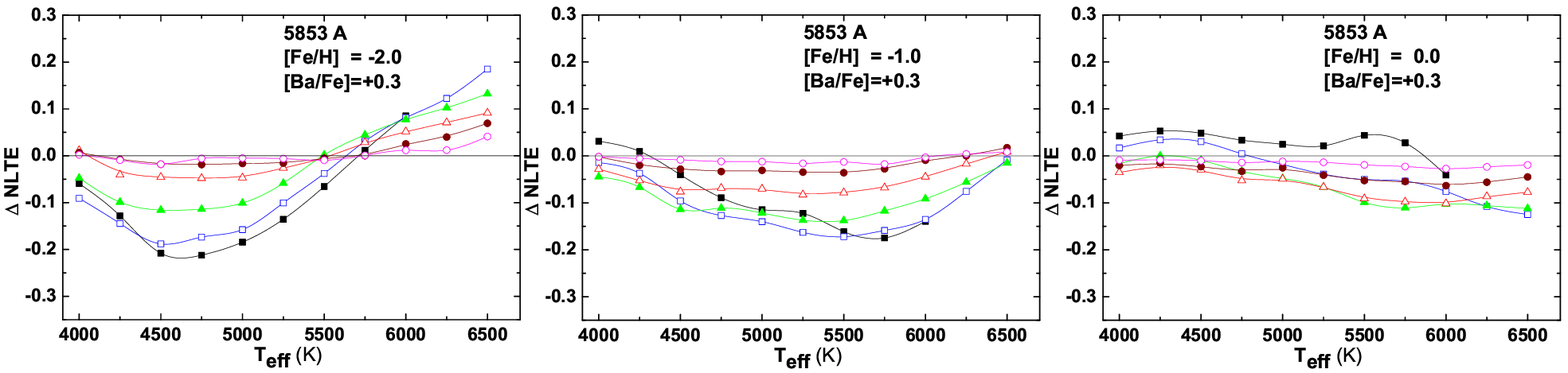}}
\resizebox{\hsize}{!}                   
{\includegraphics {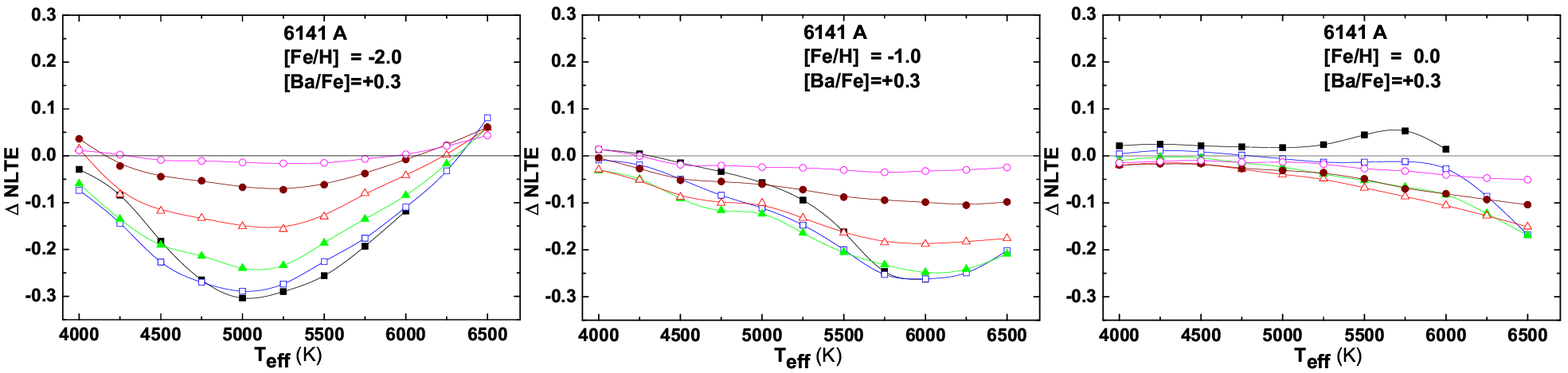}}
\resizebox{\hsize}{!}                   
{\includegraphics {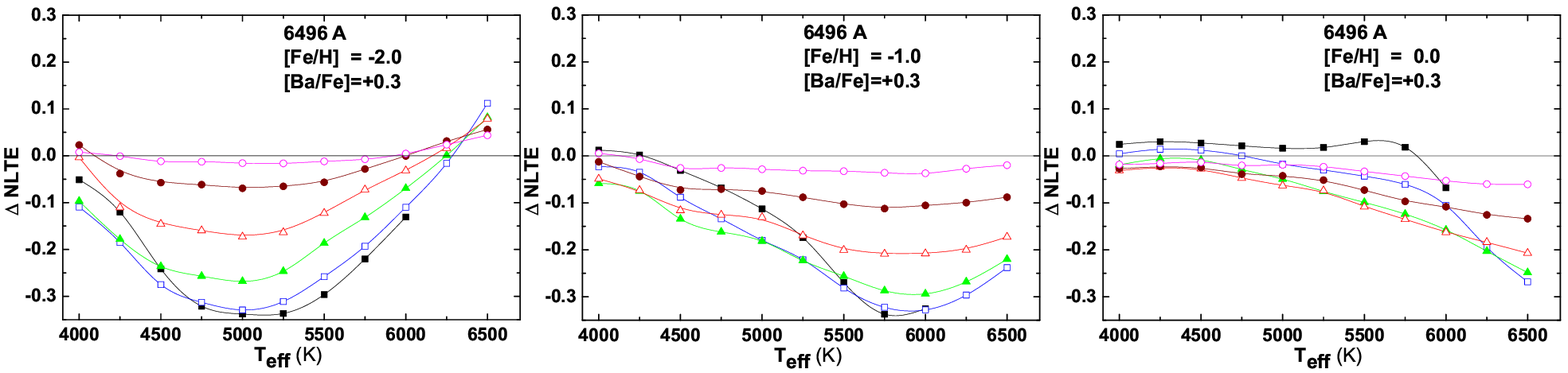}}
\caption[]{Same as Fig. \ref{dnltem3} for [Ba/Fe] = +0.3.}
\label {dnltep3}
\end{figure*}

\begin{figure*}
\resizebox{\hsize}{!}                   
{\includegraphics {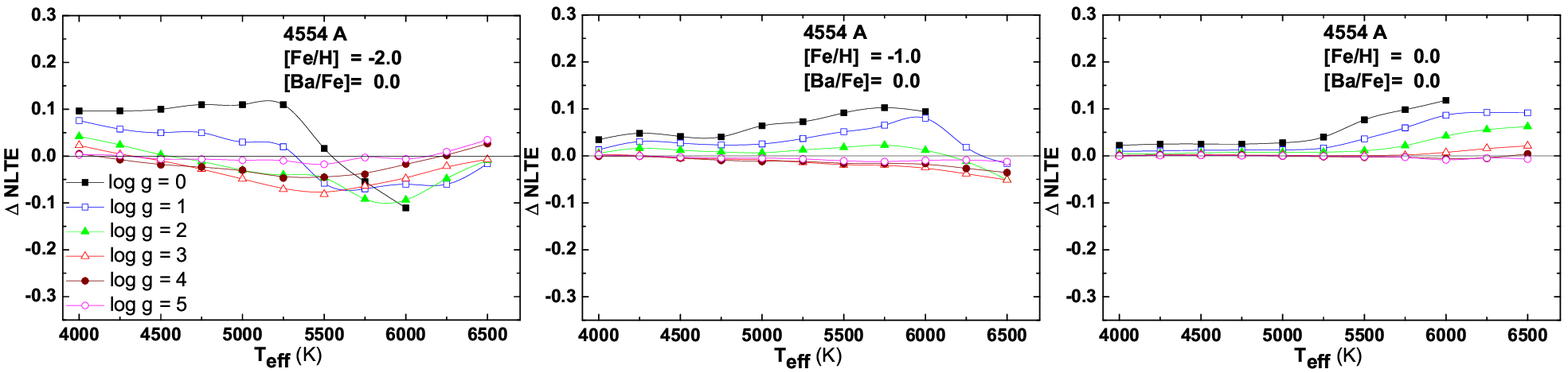}}
\resizebox{\hsize}{!}                   
{\includegraphics {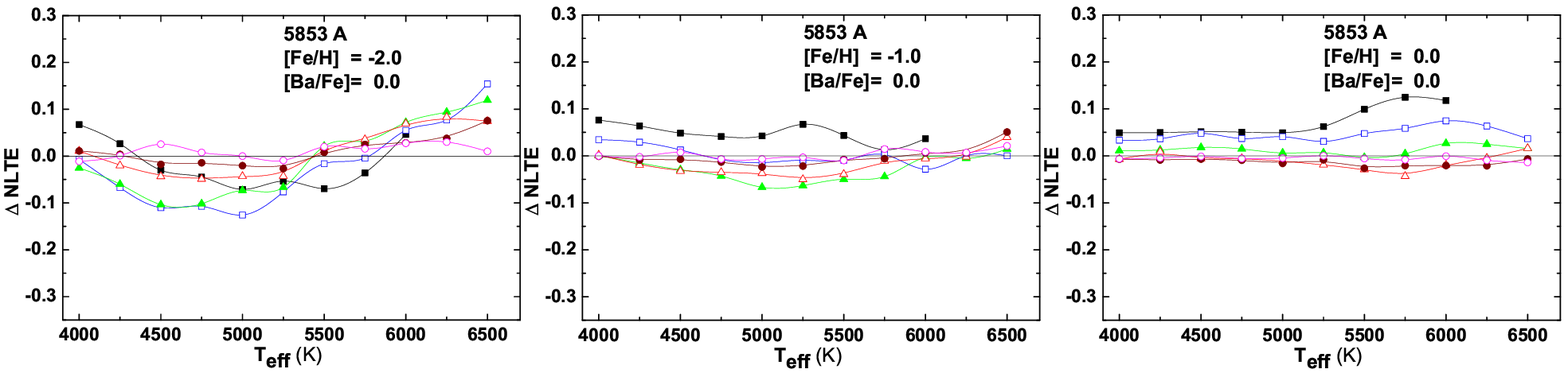}}
\resizebox{\hsize}{!}                   
{\includegraphics {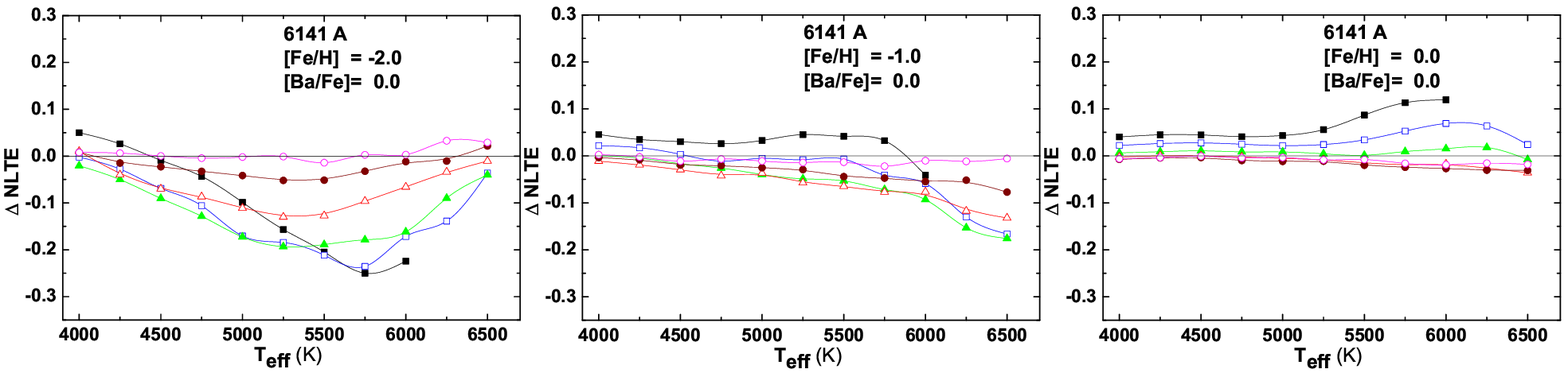}}
\resizebox{\hsize}{!}                   
{\includegraphics {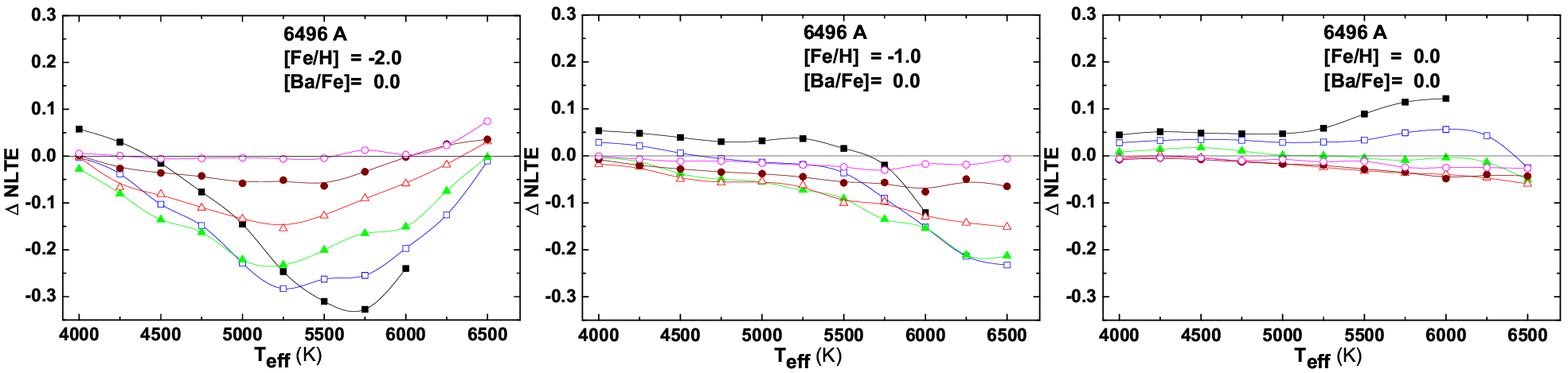}}
\caption[]{Same as Fig. \ref{dnltem3}, but for [Ba/Fe] = 0 and 
a microturbulent velocity of 0 km~s$^{-1}$.}
\label {dnltev0}
\end{figure*}

The determined LTE barium abundances can be corrected using our NLTE 
corrections listed in the grid. In Figs. \ref{dnltem3} - \ref{dnltev0} we 
show how these corrections depend upon the effective temperature, gravity, 
metallicity, and relative barium-to-iron abundance. An idea on how the 
corrections depend on the microturbulent velocity  can be obtained 
by comparing Figs.~\ref{dnltep0} and \ref{dnltev0}. 

All these figures can give a preliminary idea about the necessity 
to apply corrections for a given range of the stellar fundamental 
parameters.

\section{Conclusion}

Barium is a key element in the context of  stellar evolution
and Galactic chemical evolution.
The determination of its abundance has implications
on the evolution of AGB
stars, on the origin of heavy elements in globular clusters, on the role of
spinstars, and 
on supernovae ejecta. Therefore, it is important to derive
its abundance in stars in a reliable way that is based on the modern NLTE 
approach.

We presented a grid of NLTE equivalent widths of the \ion{Ba}{ii} 
lines 4554, 5853, 6141, and 6496 \AA~ that is intended for the barium 
abundance determination in cool dwarf, giant, and supergiant 
stars of metallicities from --2.0 to +0.5. The bluest 4554\,\AA~line is 
hard to fit in LTE, and the NLTE grid-abundance for this line deviates 
by up to 0.46\,dex compared to its NLTE profile fitting. Our tests showed 
that this line is not suitable for accurate abundance analyses in stars with 
metallicities higher than [Fe/H]$=-1$. By comparing LTE and NLTE abundances, 
it can be seen that the LTE fits for the weaker 5853\,\AA~line tend to 
yield LTE abundances close to the NLTE ones, and that the difference 
between the LTE and NLTE abundance for three red Ba lines is on average $\pm0.1$\,dex. However, in several cases, even at the higher metallicities, 
the $\Delta$NLTE is larger than this value, and NLTE corrections should be 
applied. Such corrections are also included in the grid. We showed that the 
most reliable NLTE grid-based barium abundances can be derived using the 
equivalent widths of three lines: 5853, 6141, and 6496\,\AA~. We plan to use 
this grid in the Gaia-ESO spectroscopic survey. However, it can generally 
also be used by spetroscopists working with elemental abundance analysis, 
no matter how large the samples are that they work with.

\begin{acknowledgements}
SAK and SMA acknowledge the SCOPES grant No. IZ73Z0-152485 for financial 
support. SMA is thankful to the GEPI Department and Paris Observatoire 
administration for their hospitality during his visit and to CNRS for 
the financial support. C.J.H. was supported by a research grant 
(VKR023371) from VILLUM FONDEN. E.C. is grateful to the FONDATION MERAC 
for funding her fellowship. The authors are thankful to the referee for 
their comments.

\end{acknowledgements}


\begin{thebibliography}{}

\bibitem[{Andrievsky} {et al.}(2009)]{And09}
Andrievsky, S. M., Spite, M., Korotin, S. A., et al.
2009, \aap, 494, 1083

\bibitem[{Andrievsky} {et al.}(2010)]{And10}
Andrievsky, S. M., Kovtyukh, V. V., Wallerstein, G., Korotin, 
S. A., \& Huang, Wenjin, 2010, \pasp, 122, 877

\bibitem[{Andrievsky} {et al.}(2013)]{And13}
Andrievsky, S. M., L\'epine, J. R. D., Korotin, S. A., et al., 
2013, \mnras, 428, 3252

\bibitem[{Andrievsky} {et al.}(2014)]{And14}
Andrievsky, S. M., Luck, R. E., \& Korotin, S. A., 
2014, \mnras, 437, 2106 

\bibitem[{Bisterzo} {et al.}(2010)]{Bis10}
Bisterzo, S., Gallino, R., Straniero, O., Cristallo, S., \& Kappeler, F., 
2010, \mnras, 404, 1529

\bibitem[{Britavskiy} {et al.}(2010)]{Bri10}
Britavskiy, N. E., Andrievsky, S. M., Korotin, S. A., \& Martin, P.,
2010, \aap, 519, 74
 
\bibitem[{Britavskiy} {et al.}(2012)]{Bri12}
Britavskiy, N. E., Andrievsky, S. M., Tsymbal, V. V., et al., 
2012, \aap, 542, 104

\bibitem[{Busso} {et al.}(1999)]{Bus99}
Busso, M., Gallino, R., \& Wasserburg, G. J., 1999, \araa, 37, 239

\bibitem[{Cameron} {}(1982)]{Cam82}
Cameron, A. G. W., 1982, \apss, 82, 123

\bibitem[{Carlsson} {}(1986)]{Car86}
Carlsson, M., 1986, Uppsala Obs. Rep. 33

\bibitem[{Castelli} \& {Kurucz}(2003)]{CK03}
Castelli,~F., \& Kurucz,~R. L.,
2003, in: 'Modeling of Stellar Atmospheres', Proc. IAU Symp. 210, eds. N.E.~Piskunov, 
W.W.~Weiss, \& D.F.~Gray, poster A20 (CD-ROM); synthetic spectra available at 
http://cfaku5.cfa.harvard.edu/grids

\bibitem[{Christlieb} {et al.}(2004)]{Chr04}
Christlieb, N., Beers, T. C., Barklem, P. S., et al., 
2004, \aap, 428, 1027

\bibitem[{Cristallo} {et al.}(2011)]{Cri11}
Cristallo, S., Piersanti, L., Straniero, O., et al., 
2011, \apjs, 197, 17

\bibitem[{Dobrovolskas} {et al.}(2012)]{Dob12}
Dobrovolskas, V., Kucinskas, A., Andrievsky, S. M., et al.,
2012, \aap, 540, 128

\bibitem[{D'Orazi} {et al.}(2012)]{Dor12}
D'Orazi, V., Biazzo, K., Desidera, S., et al., 
2012, \mnras, 423, 2789

\bibitem[{Gallagher} {et al.}(2012)]{Gal12}
Gallagher, A. J., Ryan, S. G., Hosford, A., et al.,
2012, \aap, 538, 118

\bibitem[{Gilmore} {et al.}(2012)]{Gil12}
Gilmore, G., Randich, S., Asplund, M., et al., 
2012, The Messenger, 147, 25 

\bibitem[{Gustafsson} {et al.}(2008)]{Gus08}
Gustafsson, B., Edvardsson, B., Eriksson, K., et al.,
2008, \aap, 486, 951

\bibitem[{Hansen} {et al.}(2011)]{Han11}
Hansen, C. J., Nordstr\"om, B., \& Bonifacio, P., 
2011, \aap, 527, 65

\bibitem[{Hansen} {et al.}(2012)]{Han12}
Hansen, C. J., Primas, F., Hartman, H., et al., 
2012, \aap, 545, 31

\bibitem[{Jofre} {et al.}(2014)]{Jof14}
Jofre, P., Heiter, U., Soubiran, C., et al., 
2014, \aap, 564, 133

\bibitem[{Karakas} \& {Lattanzio}(2014)]{Kar14}
Karakas \& Lattanzio, 2014, \pasa, 31, 30

\bibitem[{Klochkova} {et al.}(2011a)]{Klo11a}
Klochkova, V., Mishenina, T., Korotin, S., et al., 
2011a, \apss, 335, 141

\bibitem[{Klochkova} {et al.}(2011b)]{Klo11b}
Klochkova, V. G., Mishenina, T. V., Panchuk, V. E., et al.,
2011b, AstBul, 66, 28

\bibitem[{Korotin} {et al.}(1999)]{Kor99}
Korotin S.A., Andrievsky S.M., Luck R.E., 
1999, \aap, 351, 168

\bibitem[{Korotin} {et al.}(2010)]{Kor10}
Korotin, S., Mishenina, T., Gorbaneva, T., \& Soubiran, C., 
2010, nuco.cnfE.100

\bibitem[{Korotin} {et al.}(2011)]{Kor11}
Korotin, S., Mishenina, T., Gorbaneva, T., \& Soubiran, C.,
2011, \mnras, 415, 2093

\bibitem[{Kurucz} {}(1992)]{Kur92}
Kurucz R.L., 1992, The Stellar Population of Galaxies, ed. B. Barbuy, A.
Renzini, IAU Symp. 149, 225

\bibitem[{Mashonkina} {et al.}(1999)]{Mas99}
Mashonkina, L., Gehren, T., \& Bikmaev, I., 
1999, \aap, 343, 519

\bibitem[{Mishenina} {et al.}(2009)]{Mis09}
Mishenina, T. V., Kucinskas, A., Andrievsky, S. M., et al.,
2009, BaltA, 18, 193

\bibitem[{Mishenina} {et al.}(2012)]{Mis12}
Mishenina, T. V., Soubiran, C., Korotin, S. A., Gorbaneva, T. I., 
\& Basak, N. Yu., 2012, EPJWC 1905006

\bibitem[{Mishenina} {et al.}(2013a)]{Mis13a}
Mishenina, T., Korotin, S., Carraro, G., Kovtyukh, V. V., 
\& Yegorova, I. A., 2013a, MNRAS 433, 1436

\bibitem[{Mishenina} {et al.}(2013b)]{Mis13b}
Mishenina, T. V., Korotin, S. A., Yegorova, I. A., Kovtukh, V. V., 
\& Carraro, G., 2013b, BCrAO 109, 32 

\bibitem[{Mishenina} {et al.}(2013c)]{Mis13c}
Mishenina, T. V., Pignatari, M., Korotin, S. A., et al.,
2013c, \aap, 552, 128
 
\bibitem[{Mishenina} {et al.}(2014a)]{Mis14a}
Mishenina, T. V., Kovtyukh, V. V., Yegorova, I. A., Korotin, S. A., 
\& Carraro, G., 2014a, \memsai, 85, 295 
 
\bibitem[{Mishenina} {et al.}(2014b)]{Mis14b}
Mishenina, T., Pignatari, M., Carraro, G., et al.
2014b, arXiv1411.1422

\bibitem[{Rutten} {}(1978)]{Rut78}
Rutten, R.J., 1978, \solphys, 56, 237

\bibitem[{Sneden} {}(1973)]{Sne73}
Sneden, C., 1973, PhDT, 35, 28 

\bibitem[{Sneden} {et al.}(2014)]{Sne14}
Sneden, C., Lucatello, S., Ram, R. S., Brooke, J. A. S., \& Bernath, P., 
2014, \apjs, 214, 26

\bibitem[{Thygesen} {et al.}(2014)]{Thy14}
Thygesen, A. O., Sbordone, L., Andrievsky, S., et al.
2014, \aap, 572, 108

\end{thebibliography}
\end{document}